\PassOptionsToPackage{margin=2cm}{geometry}

\documentclass[sn-basic,hidelinks]{sn-jnl}

\usepackage{graphicx}
\usepackage{amssymb}
\usepackage{amsmath}
\usepackage[percent]{overpic}
\usepackage{color}
\usepackage{bm}
\usepackage{siunitx}
\usepackage{booktabs}
\usepackage{float}
\usepackage{subcaption}

\DeclareMathOperator*{\argmin}{arg\,min}

\newcommand{\smatrix}[2]{\left#2\begin{array}{#1}}
\newcommand{\ematrix}[1]{\end{array}\right#1}
\newcommand{\R}{\mathbb{R}} 
\def\dt{\Delta t}
\def\vphi{\varphi}

\def\zerob{\bm{0}}
\def\bb{\bm{b}}
\def\db{\bm{d}}
\def\eb{\bm{e}}
\def\nb{\bm{n}}
\def\ub{\bm{u}}
\def\vb{\bm{v}}
\def\Cb{\bm{C}}
\def\Eb{\bm{E}}
\def\dotdb{\bm{\dot d}}
\def\dotub{\bm{\dot u}}
\def\ddotub{\bm{\ddot u}}
\def\epssb{\bm{\epsilon}}
\def\sigmab{\bm{\sigma}}
\def\vepsb{\bm{\varepsilon}}
\def\nablab{\bm{\nabla}}

\def\thetab{\bm{\theta}}

\def\onebold{\mathbf{1}}

\def\bbold{\mathbf{b}}
\def\pbold{\mathbf{p}}
\def\ubold{\mathbf{u}}

\def\ybold{\mathbf{y}}
\def\Bbold{\mathbf{B}}
\def\Cbold{\mathbf{C}}
\def\Dbold{\mathbf{D}}

\def\Fbold{\mathbf{F}}
\def\Gbold{\mathbf{G}}
\def\Kbold{\mathbf{K}}
\def\Mbold{\mathbf{M}}
\def\Nbold{\mathbf{N}}
\def\ddotubold{\mathbf{\ddot u}}
\def\dotubold{\mathbf{\dot u}}
\def\dotpbold{\mathbf{\dot p}}

\def\dotu{{\dot u}}

\def\pder#1#2{\frac{\partial #1}{\partial #2}}

\def\Point{\bm{x}}

\def\NumTimeSteps{N+1}

\def\NodalDisplacementsAtTcurrent{\ubold^n}
\def\NodalVelocitiesAtTcurrent{\dotubold^n}
\def\NodalAccelerationsAtTcurrent{\ddotubold^n}
\def\NodalPotentialAtTcurrent{\pbold^n}
\def\NodalDotPotentialAtTcurrent{\dotpbold^n}
\def\SensorPotentialAtTcurrent{\bar{p}^n}

\def\PreviousTimestep{n-1}
\def\NodalDisplacementsAtTprevious{\ubold^{\PreviousTimestep}}
\def\NodalVelocitiesAtTprevious{\dotubold^{\PreviousTimestep}}
\def\NodalAccelerationsAtTprevious{\ddotubold^{\PreviousTimestep}}
\def\NodalPotentialAtTprevious{\pbold^{\PreviousTimestep}}
 \def\SensorPotentialAtTprevious{\bar{p}^{\PreviousTimestep}}
\def\SimulatedVibrometerVelocity{\dotu_z^L}
\def\SimulatedVibrometerVelocityAtTmeas{\SimulatedVibrometerVelocity(t_i)}
\def\SimulatedPZSensorVoltage{\bar{p}}
\def\SimulatedPZSensorVoltageAtTmeas{\SimulatedPZSensorVoltage(t_i)}

\def\TimeStepSet{\mathcal{T}}
\def\MeasurementTimesSet{\mathcal{T}_m}
\def\NumMeasurements{N_m}

\def\ParameterVector{\thetab}
\def\Parameter{\theta}

\def\MeasurementData{\mathbf{Y}}

\def\NormalizationVelocity{s_{v_z}}
\def\NormalizationVoltage{s_V}

\begin{document}

\title[A Cantilever Beam with a Piezoelectric Disc Sensor]{Experimental and Numerical Study of the Transient Response of a Cantilever Beam with a~Piezoelectric Disc Sensor}

\author*[1,2]{Radek Kolman}\email{kolman@it.cas.cz}
\author[1,3]{Robert Cimrman}\email{cimrman3@ntc.zcu.cz}
\author[1]{Ladislav Musil}\email{musil@it.cas.cz}
\author[4]{Moritz Frey}\email{moritz.frey@unibw.de}
\author[1]{Jarom\'{\i}r Kylar}\email{kylar@it.cas.cz}
\author[4]{Sebastian Brandstaeter}\email{sebastian.brandstaeter@unibw.de}
\author[1]{Vojt\v{e}ch Kotek}
\author[4]{Alexander Popp}\email{alexander.popp@unibw.de}
\author[1]{Jan Kober}\email{kober@it.cas.cz}

\affil*[1]{Institute of Thermomechanics, v.v.i., The Czech Academy of Sciences, Dolej\v{s}kova 1402/5; 182 00, Prague 8, Czech Republic}
\affil[2]{The College of Polytechnics Jihlava, Tolstého 16, 586 01 Jihlava, Czech Republic}
\affil[3]{New Technologies - Research Centre and Faculty of Applied Sciences, University of West Bohemia, Pilsen, Czech Republic}
\affil[4]{Institute for Mathematics and Computer-Based Simulation, University of the Bundeswehr Munich, Werner-Heisenberg-Weg 39, 85577 Neubiberg, Germany}

\abstract{
Online and real-time sensing and monitoring of the health state of complex structures, such as aircraft and critical components of power stations, are essential aspects of research in dynamics. Several types of sensors are used to capture dynamic responses and monitor changes during the operation of critical parts of complex systems. Piezoelectric (PZ) materials belong to a class of electroactive materials that convert mechanical deformation into an electrical response. For example, PZ ceramics or PVDF foils are employed for online sensing of the time history of mechanical deformation.
This paper focuses on the dynamical response of a cantilever beam structure equipped with a glued PZ sensor and combines experimental and modelling approaches to achieve accurate and reliable results. The time history of the normal velocity at a point on the beam surface was recorded with a laser vibrometer during transient vibrations of the beam, triggered by the sudden removal of a mass load at the beam’s free end. Simultaneously, the output voltage of the PZ sensor was measured with an electronic device.
An elastodynamic model of a cantilever beam coupled with a piezoelectric sensor is introduced, along with its discretization using the finite element method. The mathematical model includes additional terms that enforce a floating-potential boundary condition to maintain a constant charge on one of the sensor’s electrodes and is presented in an extended form suitable for sensitivity analysis or parameter identification.
The model implementation is validated using a numerical example corresponding to the experimental setup. The computed results show good agreement with the experimental data. Furthermore, values of the Rayleigh damping parameters were identified based on the experimental measurements.
}


\maketitle
\newpage
\section{Introduction}
The structural integrity and operational reliability of complex engineering systems, such as aircraft and critical components of power stations, are of paramount importance for both safety and efficiency. Failures in these structures can lead to catastrophic consequences, including loss of life, economic damage, and environmental disasters. Consequently, online and real-time sensing and monitoring of the health state of these structures have become essential aspects of research in structural dynamics and nondestructive testing (NDT). With advancements in smart materials, sensing technologies, and computational modeling, the field is undergoing a significant transformation toward predictive maintenance and condition-based monitoring.
In contrast, real-time structural health monitoring (SHM) aims to continuously assess the structural condition during operation, thereby enabling early fault detection, damage localization, and estimation of the remaining useful life \cite{FarrarWorden2007}. Recent advancements in NDT for SHM are presented in~\cite{Kot2021}.
The success of online SHM relies on the development and integration of robust, accurate, and sensitive sensor systems capable of capturing relevant physical phenomena such as strain, stress, displacement, vibration, and acoustic emissions. Several sensing technologies are currently used in practice \cite{Hassani2023}.

Often, smart materials are employed to sense mechanical responses, see~\cite{Bengisu2018}.
Smart materials, also referred to as intelligent or responsive materials, are engineered materials whose properties can be significantly altered in a controlled manner by external stimuli such as stress, electric or magnetic fields, and temperature. Examples include piezoelectric (PZ) materials, shape-memory alloys, electroactive polymers, electrostrictive materials, magnetostrictive materials, and others; see \cite{Shahinpoor2020}. These materials and the structures incorporating them are used to sense dynamic events. PZ materials, in particular, are employed for sensing mechanical responses, energy harvesting, and electromechanical actuation.
PZ materials such as lead zirconate titanate (PZT) and polyvinylidene fluoride (PVDF) are widely used due to their ability to convert mechanical energy into electrical signals \cite{Bibhudutta2024}. These sensors can be embedded or surface-mounted and are capable of detecting subtle changes in strain and vibration \cite{Ferreira2022}. Their high-frequency response and small form factor make them ideal for monitoring high-speed dynamic events in aircraft wings, rotor blades, and turbine structures.

The standard numerical method in elastodynamics is the finite element method (FEM) \cite{zienkiewicz_finite_2005}. The piezoelectric effect can also be incorporated into numerical simulations. The first publication on this topic was authored by Alik \cite{Alik1970}, who derived the matrix formulation and proposed the numerical strategy. Benjeddou \cite{benjeddou2000advances} reviewed advances in formulations and applications of finite element modeling of adaptive structural elements.


Beyond sensing, a modern SHM system requires integration with high-fidelity computational models to interpret measured data and predict future behavior. Finite element models are commonly employed to simulate the structural response under various loading conditions. When coupled with sensor inputs, these models can serve as the foundation for digital twins—real-time virtual replicas of physical systems that update dynamically based on operational data.





In this paper, we study the dynamic response of a steel beam with a PZ sensor glued on top, both experimentally and computationally. The experimental setup is described in Section~\ref{sec:problem}, and the measurement results are presented in Section~\ref{sec:experiments}. The measured first eigenfrequency is supported by an analytical model in Section~\ref{sec:analytical}. The mathematical model introduced in Section~\ref{sec:model} includes additional terms that enforce the floating-potential boundary condition to maintain constant charge on one of the sensor electrodes and is presented in an extended form suitable for sensitivity analysis or parameter identification. This capability is demonstrated using a numerical example corresponding to the experimental setup, where the mechanical material parameter and Rayleigh damping coefficients are identified, yielding good agreement with measurements. The calculations employ the open-source multiscale FEM solver SfePy \cite{Cimrman_Lukes_Rohan_2019}, which is used for both direct and inverse problem computations. This study contributes to ongoing work on the dynamics of metamaterials made of 3D-printed structures, including PZ actuators and sensors.

\newpage
\section{Problem definition}
\label{sec:problem}
\begin{figure}[htp!]
  \centering
\includegraphics[angle=270, width=0.55\linewidth]{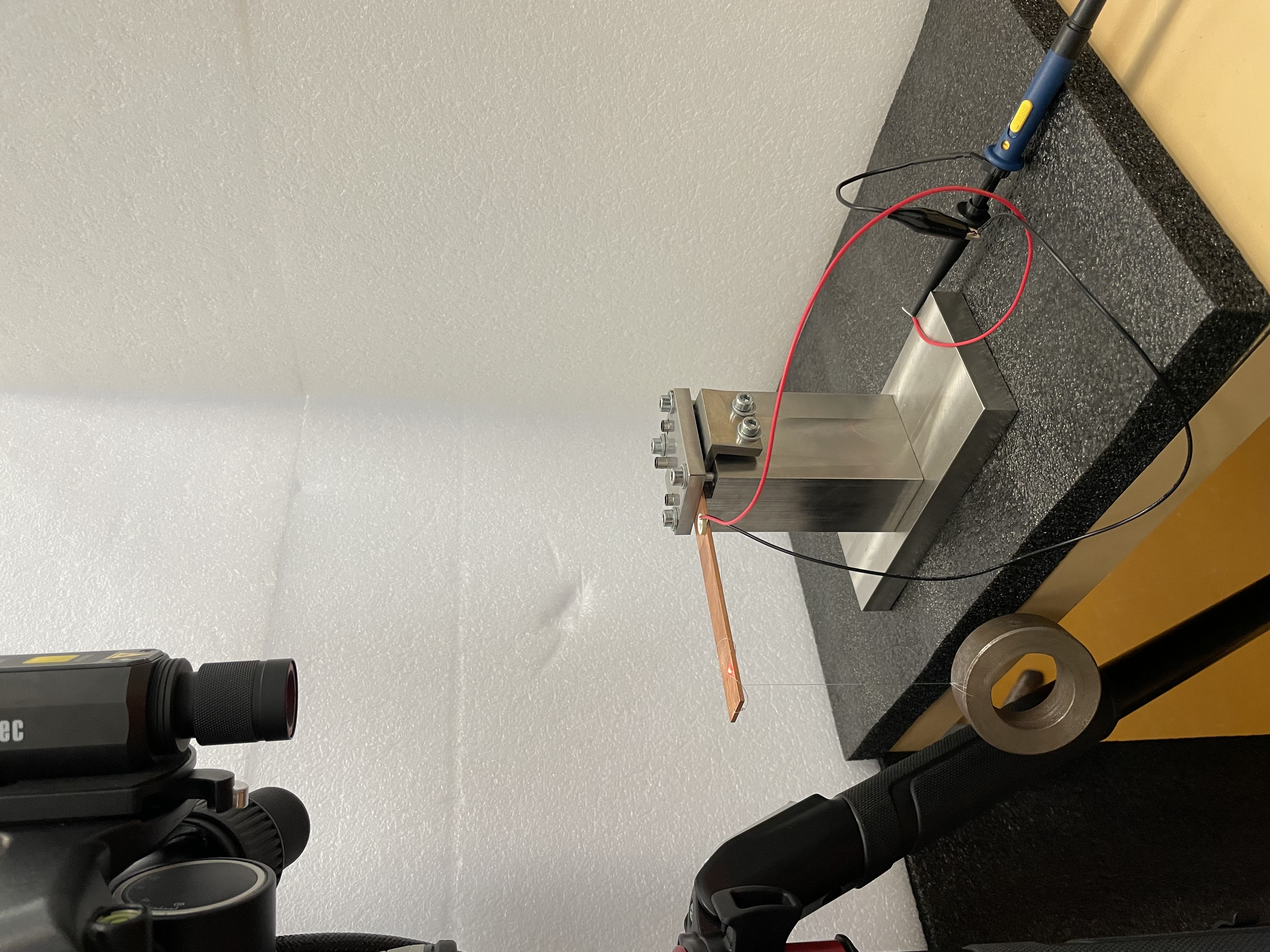}
\caption{The experimental setup for measuring the dynamic response of a beam structure subjected to a suddenly interrupted load caused by the removal of a mass attached to the beam end. The beam structure is clamped to a robust holder.}
\label{fig:Experimental_setup}
\end{figure}

We study the dynamic response of a steel cantilever beam subjected to a suddenly removed load applied by a mass attached to the end of the beam. The experimental setup is shown in Fig.~\ref{fig:Experimental_setup}, where the strip-shaped beam structure is clamped to a robust holder.

\begin{figure}[htp!]
  \centering
\includegraphics[width=0.65\linewidth]{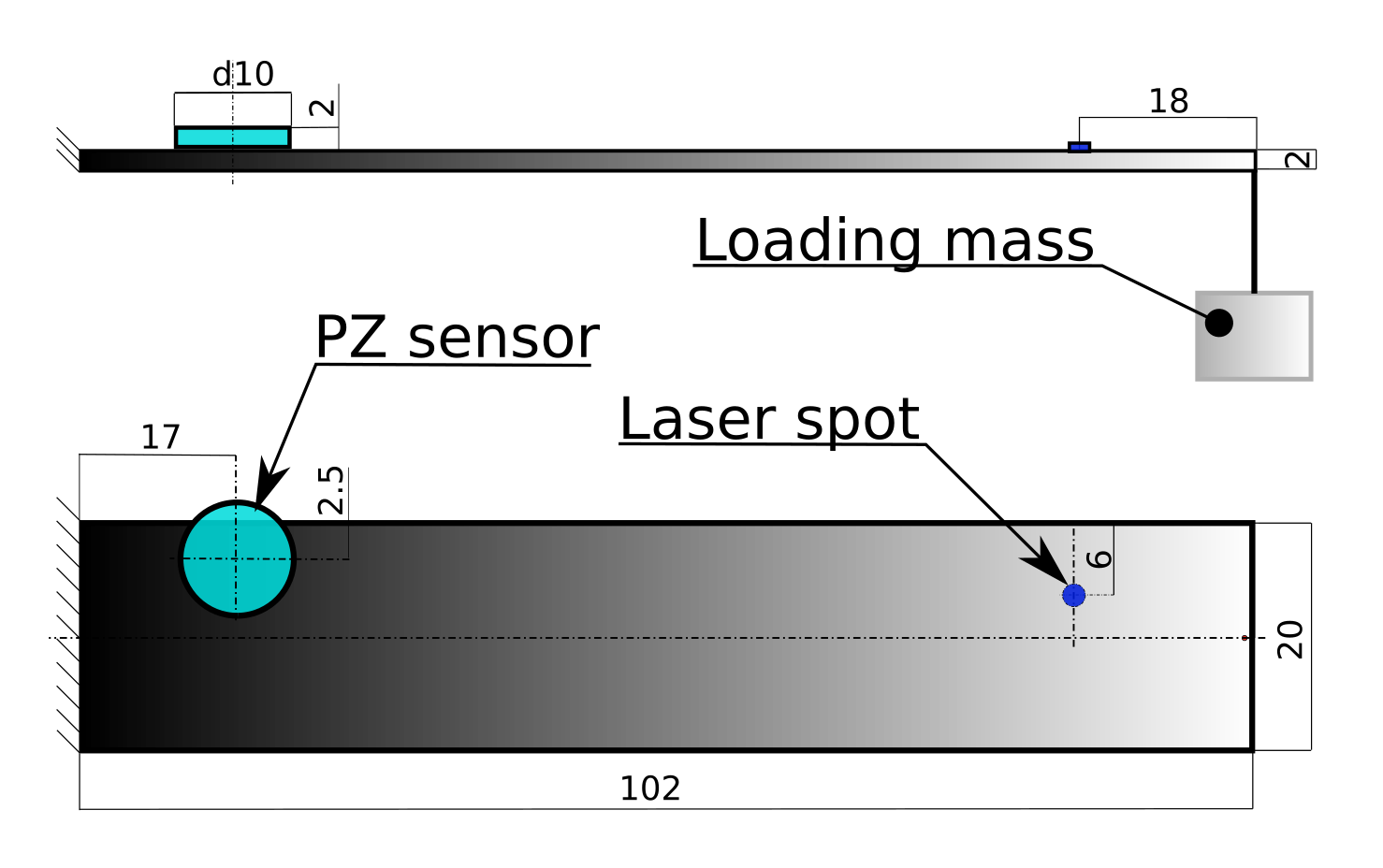}
  \caption{The schematic of the experiment, showing the indicated dimensions and the positions of the PZ sensor, laser spot, and loading mass.}
\label{fig:Scheme_of_experiment}
\end{figure}

Initially, the beam is pre-deformed due to the static force induced by the weight of an object with a mass of 0.282~kg, which is tied to the beam end by a nylon line. The force corresponding to the attached mass is $F = 2.766$~N. Subsequently, the nylon line is cut, and the beam undergoes free vibration.
The dynamic response of the beam is monitored using a PZ sensor disc and a laser vibrometer. The voltage across the electrodes of the PZ sensor is recorded, along with the normal component of the velocity at a point on the beam surface measured by the laser vibrometer. The beam dimensions, as well as the positions of the PZ sensor and the laser measurement point, are illustrated in Fig.~\ref{fig:Scheme_of_experiment}.

The beam material is mild steel, and the total mass of the steel beam is $45.803$~g. The overall length of the beam structure is $150$~mm, with a width of $b = 20$~mm and a thickness of $h = 1.905$~mm. The active length of the beam is $L = 102$~mm. Based on these values, the material density of the beam is $\rho = 8014.5~\si{kg/m^3}$.

\section{Experiments}
\label{sec:experiments}

For sensing the dynamic response of the beam, a piezoelectric sensor disc with a diameter of 10~mm and a thickness of 2~mm, made of PIC~181 material (PI Ceramic, https://www.piceramic.com/
), was used. The voltage across the Ag screen-printed electrodes of the PZ sensor was measured using a Picoscope~3404D oscilloscope. PIC~181 is a modified lead zirconate–lead titanate material characterized by an extremely high mechanical quality factor and a high Curie temperature. It is designed for high-power acoustic applications. In addition, its good temperature and time stability of dielectric and elastic constants makes it well suited for resonance-mode ultrasonic applications, and it has proven particularly successful in piezo-motor drives. Other important parameters of the sensor include a resonant frequency (thickness mode) of $10.550$~kHz, a resonant frequency (radial mode) of $227$~kHz, and an electrical capacitance of $0.435$~nF.

For velocity measurements, a Polytec CLV-2000 laser vibrometer equipped with a CLV-700 sensor head was used. The laser unit was mounted on a tripod positioned above the beam, and the laser beam was directed onto the beam at a distance of approximately $30$~cm. To ensure sufficient optical reflection, a reflective label was attached to the beam surface at the location indicated in Fig.~\ref{fig:Scheme_of_experiment}, as the untreated surface did not provide adequate reflectivity for reliable measurements. The quality of the reflected signal was monitored using the input module located on the front panel of the CLV-2000 unit.

\subsection{Measurement of the material parameters of the beam}
\label{sec:mmpb}

For the vibration analysis of the beam, its mechanical material parameters are required, namely the Young’s modulus and Poisson’s ratio used in the elastic constitutive model.

The wave speed of longitudinal waves propagating through the steel beam in the thickness direction was measured using the pulse–echo method and identified as $c_1 = 5693$~\si{m/s}. The tabulated value of Poisson’s ratio for steel, $\nu = 0.3$, was adopted. For this value, the corresponding Young’s modulus is predicted as $E = 189$~GPa, based on the relation for the longitudinal wave speed
$c_1 = \sqrt{(2G + {\Lambda}) / \rho},$ where ${\Lambda} = \frac{\nu E}{(1+\nu)(1-2\nu)}$ and $G=\frac{E}{2(1+\nu)}$, see \cite{achenbach2012wave}.



\subsection{Experimental outputs}
\label{sec:oe}

\begin{figure}[htp!]
  \centering
  \includegraphics[width=0.48\linewidth]{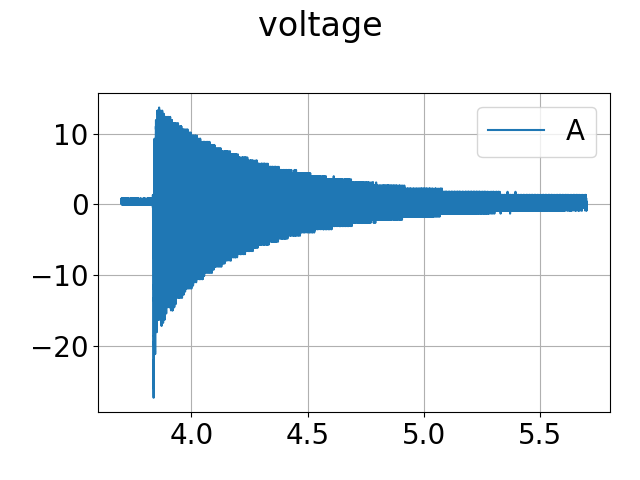}
  \includegraphics[width=0.48\linewidth]{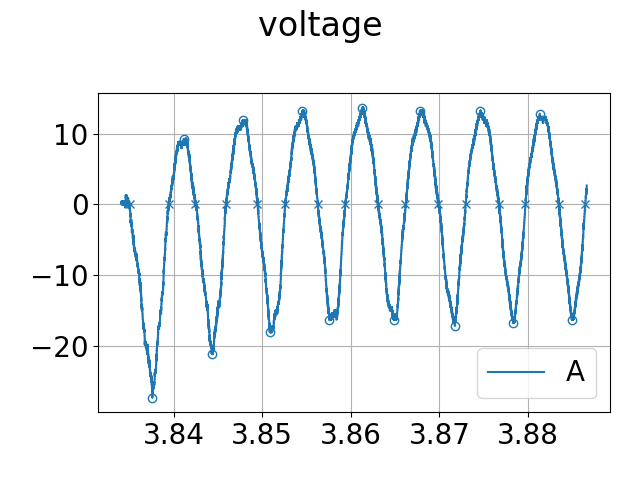}
  \caption{Time history of the voltage (in V) measured by the PZ sensor: full record (left) and zoomed view of the transient vibration onset (right).}
\label{fig:voltage_of_experiment}
\end{figure}

\begin{figure}[htp!]
  \centering
  \includegraphics[width=0.48\linewidth]{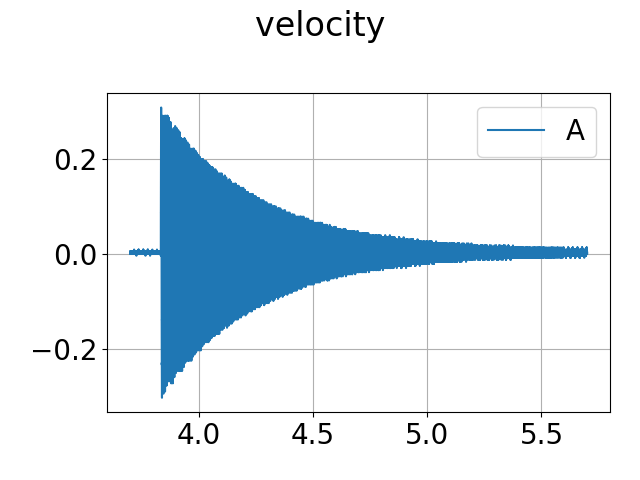}
  \includegraphics[width=0.48\linewidth]{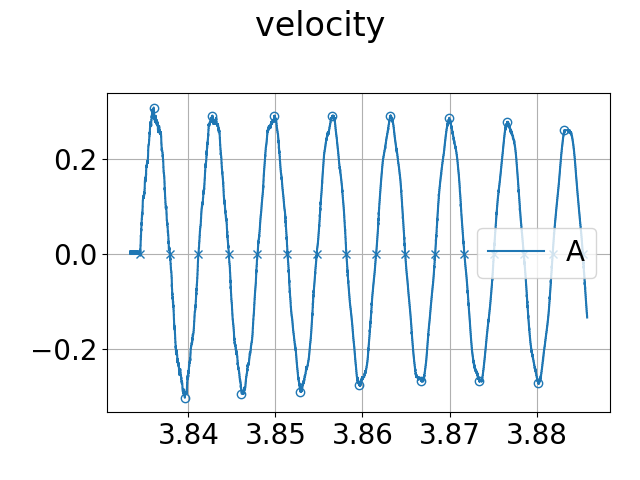}
  \caption{Time history of the normal velocity (in m/s) at the laser spot measured by the laser vibrometer: full record (left) and zoomed view of the transient vibration onset (right).}
\label{fig:velocity_of_experiment}
\end{figure}

The experiments resulted in time-history records of the following quantities. The time history of the voltage measured by the PZ sensor is shown in Fig.~\ref{fig:voltage_of_experiment} (left), while the time history of the normal velocity at the laser spot is presented in Fig.~\ref{fig:velocity_of_experiment} (left). The zoomed-in portions of the records in Fig.~\ref{fig:voltage_of_experiment} (right) and Fig.~\ref{fig:velocity_of_experiment} (right) illustrate the onset of the transient vibrations. The first eigenfrequency of the structure, $f_{1,\mathrm{measured}} = 147.59$~Hz, was identified from the experimental data using fast Fourier transform (FFT)–based signal processing in MATLAB.

\section{Analytical estimate of the first eigenfrequency of the beam}
\label{sec:analytical}
The first eigenfrequency can be estimated using an analytical model of linear, undamped vibrations of a cantilever beam based on the Bernoulli–Navier hypothesis. An analytical solution is given in Rao~\cite{rao1995mechanical}, where the first eigenfrequency is computed as
\begin{equation}\label{eq:frequencies}
f_{1}=\frac{1}{2\pi}(\beta_1 l)^2 \, \sqrt{\frac{EI}{\rho A L^4}},
\end{equation}
where, in the present case, $(\beta_1 l)=1.8751$, the second moment of area is $I=\frac{1}{12}bh^3$, the cross-sectional area is $A=bh$, the steel density is $\rho = 8014.5$~\si{kg/m^3}, and the Young’s modulus is $E = 189$~GPa. Using these values, the analytical first eigenfrequency is obtained as $f_{1,,\mathrm{analytical}}=143.62$~Hz.

\section{The finite element model}
\label{sec:model}

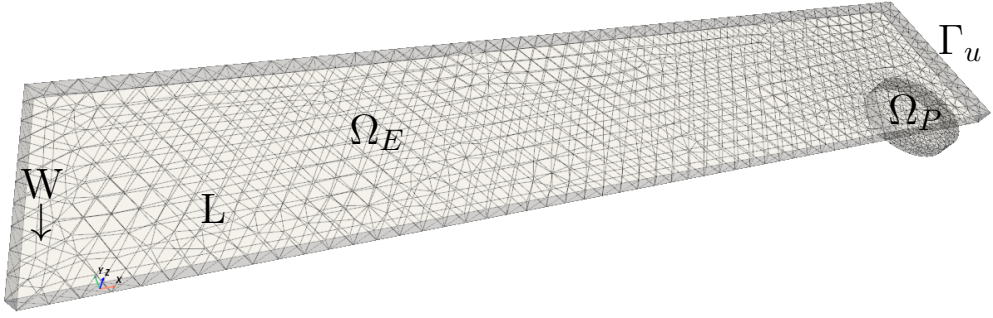
\begin{figure}[htp!]
  \centering

  \begin{overpic}[width=0.8\linewidth]{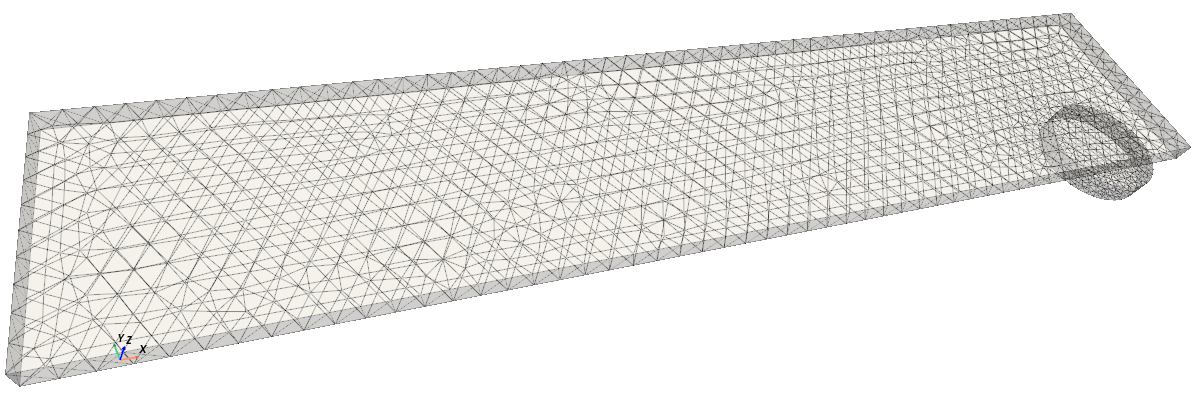}
    \put (2,12.5) {\Large W}
    \put (3,9) {\Large $\downarrow$}
    \put (20,10) {\Large L}
    \put (35,18) {\Large $\Omega_E$}
    \put (89,20) {\Large $\Omega_P$}
    \put (94,27) {\Large $\Gamma_u$}
  \end{overpic}

  \caption{The computational domain with the finite element mesh. A static weight is attached to the bottom side at point~W at time $t = 0$~s. Dynamic motion quantities are recorded at point~L, which corresponds to the location of the laser vibrometer measurements in the experiments.}
  \label{fig:domain}
\end{figure}

Let $\Omega \subset \mathbb{R}^3$ denote the computational domain, consisting of an elastic part $\Omega_E$ corresponding to the cantilever beam and a piezoelectric part $\Omega_P$ representing the cylindrical sensor attached to the beam. A point in the domain $\Omega$ is denoted by $\Point$. In the present setting, the cantilever beam is fixed along the boundary $\Gamma_u$, as shown in Fig.~\ref{fig:domain}.

Under linear assumptions, the constitutive relations of the piezoelectric solid in $\Omega_P$ can be written as
\begin{equation}
\sigmab = \Cb^P \vepsb - \eb^{T} \Eb \;,\quad
\db = \eb \vepsb + \epssb \Eb \;,\quad
\vepsb = \frac{1}{2}\bigl(\nablab \ub + \nablab^{T} \ub\bigr) \;,\quad
\Eb = - \nablab \vphi \;,
\end{equation}
where the mechanical stress $\sigmab$ (in Voigt notation) and the electric displacement $\db$ depend on the mechanical Cauchy strain $\vepsb$ and the electric field vector $\Eb$. Here, $\ub$ denotes the mechanical displacement vector, $\vphi$ the electric potential, $\Cb^P$ the matrix of elastic stiffness moduli at constant electric field, $\eb$ the matrix of piezoelectric coupling moduli, and $\epssb$ the permittivity matrix at constant strain. In the elastic domain $\Omega_E$, the constitutive relation reduces to $\sigmab = \Cb^E \vepsb$.

\subsection{Piezo-elastodynamic problem: strong form}
\label{sec:strong}

The strong form of the piezo-elastodynamic problem to be solved over
$t \in [0, T]$ is as follows: Find $\ub$, $\vphi$, and $\bar\vphi(t)$ such that
\begin{equation}
  \label{eq:pedp}
  \begin{split}
    \rho \ddotub - \nablab \cdot \sigmab &= \bb \mbox{ in }
    \Omega \times [0, T]\;, \\
    \nablab \cdot \db &= 0 \mbox{ in } \Omega_P \times [0, T]\;, \\
    \ub &= 0 \mbox{ on } \Gamma_u \times [0, T]\;, \\
    \vphi &= 0 \mbox{ on } \Gamma_{p0} \times [0, T]\;, \\
    \vphi &= \bar\vphi(t) \mbox{ on } \Gamma_{pQ} \times [0, T]
    \mbox{ such that }
    \int_{\Gamma_{pQ}} \dotdb \cdot \nb = \bar\vphi(t) / R
    +  C \dot{\bar\vphi} \;, \\
    \ub |_{t=0} &= \ub^0 \;,\quad
                  \dotub |_{t=0} = \dotub^0\;,\quad
                  \vphi|_{t=0} = \vphi^0 \mbox{ in } \Omega \;,
  \end{split}
\end{equation}
where $\rho$ denotes the material density, $\bb$ the self-weight (body) forces, and $\nb$ the outward unit normal. Moreover,
$\dotdb = \epssb \nablab \dot\vphi - \eb \vepsb(\dotub)$.

Zero electric potential is prescribed on the bottom electrode of the piezo-sensor, $\Gamma_{p0}$. The potential $\vphi$ is also required to be spatially constant on the top electrode, $\Gamma_{pQ}$, but its value evolves in time and is represented by the additional unknown $\bar\vphi(t)$. This unknown is determined such that the electric current through $\Gamma_{pQ}$ matches the current in the external circuit characterized by the oscilloscope resistance $R$ and capacitance $C$. Homogeneous Neumann boundary conditions are applied on the remaining parts of $\partial \Omega$ not specified above. The initial charge state is determined from the initial conditions $\ub^0$, $\dotub^0$, and $\vphi^0$, which correspond to a static equilibrium under self-weight and a localized surface force due to the weight attached at point~$W$ in Fig.~\ref{fig:domain}.

\subsection{Piezo-elastodynamic problem: weak form}
\label{sec:weak}

Next, we define the weak formulation of the mathematical model \eqref{eq:pedp}, including Rayleigh (proportional) damping. Since the value of the Dirichlet boundary condition for the electric potential on $\Gamma_{pQ}$ is unknown, this condition must be enforced weakly. The weak formulation is then stated as follows.
Let
$V_0^u(\Omega) = \{\ub \in [H^1(\Omega)]^3, \ub = \zerob \mbox{ on }
\Gamma_u\}$,
$V_0^\vphi(\Omega_P) = \{\vphi \in H^1(\Omega_P), \vphi = 0 \mbox{ on }
\Gamma_{p0}\}$. We seek $\ub(t)$, $\vphi(t)$, $\bar\vphi(t)$ such that
\small
\begin{align}
  \int_\Omega \rho \vb \cdot \ddotub
  + \int_\Omega \vepsb(\vb)^T \Cb \vepsb(\ub)
  + \alpha \int_\Omega \rho \vb \cdot \dotub
  + \beta \int_\Omega \vepsb(\vb)^T \Cb \vepsb(\dotub) \nonumber \\
  - \int_\Omega \vb \cdot \bb
  + \int_{\Omega_P} \vepsb(\vb)^T \eb^T \nablab \vphi
  - \int_{\Gamma_{pQ}} (\eb \vepsb(\vb)) \cdot \nb (\vphi - \bar\vphi)
  & = 0 & \forall \vb \in V_0^u(\Omega) \;, \label{eq:wpedp:a} \\
  - \int_{\Omega_P} (\nablab \psi)^T \eb  \vepsb(\ub)
  + \int_{\Omega_P} (\nablab \psi)^T \epssb \nablab \vphi
  - \int_{\Gamma_{pQ}} (\epssb \nablab \vphi) \cdot \nb \psi \nonumber \\
  + \int_{\Gamma_{pQ}} (\eb \vepsb(\ub)) \cdot \nb \psi
  + \int_{\Gamma_{pQ}} (\epssb \nablab \psi) \cdot \nb (\vphi - \bar\vphi)
  + \int_{\Gamma_{pQ}} \gamma \psi (\vphi - \bar\vphi)
  & = 0 & \forall \psi \in V_0^\vphi(\Omega_P) \;, \label{eq:wpedp:b} \\
  \int_{\Gamma_{pQ}} (\epssb \nablab \dot \vphi) \cdot \nb
  - \int_{\Gamma_{pQ}} (\eb \vepsb(\dotub)) \cdot \nb
  - \bar\vphi / R - C \dot{\bar\vphi}
  & = 0  \;, \label{eq:wpedp:c} \\
  \ub & = 0 & \mbox{ on } \Gamma_u \times [0, T]\;, \\
  \vphi & = 0 & \mbox{ on } \Gamma_{p0} \times [0, T]\;, \\
  \ub(0) = \ub^0,\ \dotub(0) = 0,\ \vphi(0) &
  = \vphi^0 & \mbox{ in } \Omega \;,
\end{align}
where $\alpha$ and $\beta$ are the proportional damping coefficients, and $\gamma$ is a penalty parameter. The external circuit is modeled by \eqref{eq:wpedp:c}. The last term in \eqref{eq:wpedp:a} and the last four terms in \eqref{eq:wpedp:b} correspond to the weak enforcement of the Dirichlet boundary condition $\vphi = \bar\vphi(t)$ on $\Gamma_{pQ}$ using a non-symmetric variant of Nitsche’s method \cite{Nitsche_1971}. Unlike in electrostatics, where $\db = -\epssb \nablab \vphi$, the present piezoelectric setting yields $\db = \eb \vepsb - \epssb \nablab \vphi$, which introduces additional terms in the Nitsche-type penalization of the charge flux.

\subsection{Finite element discretization}
\label{sec:fe}

The finite element discretization of \eqref{eq:wpedp:a}–\eqref{eq:wpedp:c} is based on the approximations
\begin{equation}
  \label{eq:fea}
  \ub(\Point,t) = \Nbold_u(\Point) \ubold(t) \;, \quad
  \vphi(\Point,t) = \Nbold_\vphi(\Point) \pbold(t) \;, \quad
  \bar\vphi(t) = 1 \bar p(t) \;,
\end{equation}
where $\Nbold_u(\Point)$ and $\Nbold_\vphi(\Point)$ denote the displacement and electric potential shape function matrices, respectively. Here, $\ubold(t)$ is the vector of time-continuous nodal displacements, $\pbold(t)$ the vector of time-continuous nodal electric potentials, and $\bar p(t)$ the time-continuous electric potential at the PZ sensor.
The implicit second-order Newmark method \cite{Newmark_1959} is employed to discretize the dynamic equation \eqref{eq:wpedp:a}, while the external circuit equation \eqref{eq:wpedp:c} is discretized using a central difference scheme. Since a sensitivity analysis of the time-dependent system with respect to several parameters is intended, the discrete equations are presented in an extended form.
A uniform time discretization with time step size $\dt$ is introduced such that the discrete times are $t_n = n\dt \in [0,T]$, with $n \in \{0,1,\ldots,N\}$ and $t_N = T$. For later reference, we define the set of all discrete time instants as $\TimeStepSet = \{t_0, t_1, \ldots, t_N\}$.
The unknown quantities at time step $n \geq 1$ are
$\NodalDisplacementsAtTcurrent$,
$\NodalVelocitiesAtTcurrent$,
$\NodalAccelerationsAtTcurrent$,
$\NodalPotentialAtTcurrent$,
$\NodalDotPotentialAtTcurrent$,
$\SensorPotentialAtTcurrent$,
where, for example, $\NodalDisplacementsAtTcurrent = \ubold(t_n)$. The corresponding known quantities from the previous time step $n-1$ are denoted by the superscript $^{\PreviousTimestep}$ (again for $n \geq 1$).
Finally, the fully discrete equations in space and time, written in extended form, read as follows:
\begin{align}
  \Mbold \NodalAccelerationsAtTcurrent + \Cbold \NodalVelocitiesAtTcurrent + \Kbold \NodalDisplacementsAtTcurrent + \Bbold^T
  \NodalPotentialAtTcurrent - \Gbold^T (\NodalPotentialAtTcurrent - \onebold \SensorPotentialAtTcurrent)
  - \bbold & = 0 \;, \label{eq:dsas:a} \\
  - \Bbold \NodalDisplacementsAtTcurrent + \Gbold \NodalDisplacementsAtTcurrent + \Dbold \NodalPotentialAtTcurrent - \Fbold \NodalPotentialAtTcurrent
  + \Fbold^T (\NodalPotentialAtTcurrent - \onebold \SensorPotentialAtTcurrent)
  + \Mbold_p (\NodalPotentialAtTcurrent - \onebold \SensorPotentialAtTcurrent)
  & = 0 \;, \\
  \onebold^T \Fbold \NodalDotPotentialAtTcurrent - \onebold^T \Gbold \NodalVelocitiesAtTcurrent
  - \frac{1}{2R} (\SensorPotentialAtTcurrent + \SensorPotentialAtTprevious)
  - C \frac{\SensorPotentialAtTcurrent - \SensorPotentialAtTprevious}{\dt} & = 0 \;, \label{eq:dsas:c} \\
  \NodalVelocitiesAtTcurrent - \NodalVelocitiesAtTprevious - (1 - \gamma_N) \dt \NodalAccelerationsAtTprevious -
  \gamma_N \dt \NodalAccelerationsAtTcurrent & = 0 \;, \label{eq:dsas:d} \\
  \NodalDisplacementsAtTcurrent - \NodalDisplacementsAtTprevious - \dt \NodalVelocitiesAtTprevious - (\frac{1}{2} - \beta_N)
  \dt^2 \NodalAccelerationsAtTprevious - \beta_N \dt^2 \NodalAccelerationsAtTcurrent & = 0 \;, \label{eq:dsas:e} \\
  \NodalDotPotentialAtTcurrent + \frac{1}{\dt} \NodalPotentialAtTprevious - \frac{1}{\dt} \NodalPotentialAtTcurrent
  & = 0 \;, \label{eq:dsas:f}
\end{align}
where $\Mbold$ denotes the mass matrix, $\Cbold$ the damping matrix, $\Kbold$ the stiffness matrix, $\Bbold$ the piezoelectric coupling matrix, $\Dbold$ the electrostatic potential matrix, and $\bbold$ the vector of volume forces. The surface flux matrices
$\Fbold = \int_{\Gamma_{pQ}} \Nbold_\vphi^T \nb \epssb \Nbold_\vphi$ and
$\Gbold = \int_{\Gamma_{pQ}} \Nbold_\vphi^T \nb \eb \Nbold_{\vepsb}$ are used to weakly impose the Dirichlet boundary condition $\vphi = \bar\vphi$ on $\Gamma_{pQ}$. The matrix $\onebold$, consisting of ones, is employed to sum the rows of $\Fbold$ and $\Gbold$, thereby performing the integration appearing in \eqref{eq:wpedp:c}. Furthermore, $\Mbold_p$ denotes the potential penalty (mass) matrix.
The Newmark time integration scheme is defined by \eqref{eq:dsas:d} and \eqref{eq:dsas:e}, with parameters $\beta_N$ and $\gamma_N$, while the central difference scheme is given by \eqref{eq:dsas:f} and the averaging operator in the third term of \eqref{eq:dsas:c}. When solving the direct problem, the quantities $\NodalVelocitiesAtTcurrent$, $\NodalDisplacementsAtTcurrent$, and $\NodalDotPotentialAtTcurrent$, as expressed by \eqref{eq:dsas:d}–\eqref{eq:dsas:f}, are substituted into \eqref{eq:dsas:a}–\eqref{eq:dsas:c}. This results in three primary unknowns at each time step—namely $\NodalAccelerationsAtTcurrent$, $\NodalPotentialAtTcurrent$, and $\SensorPotentialAtTcurrent$—which are obtained by solving a linear system.

\subsection{Parameter identification}
\label{sec:pi}
In this section, we present the general framework for the parameter identification problem, which is used to determine unknown parameter values from experimental data.
To identify the parameters, measured time series are fitted to the corresponding outputs obtained from numerical simulations.
The following pairs of measurement data and simulation outputs are used:
\renewcommand{\labelenumi}{\alph{enumi})}
\begin{enumerate}
\item the velocity component $v_z(t_i)$
measured by a laser vibrometer at the point $L$ and the corresponding
simulated quantity $\SimulatedVibrometerVelocityAtTmeas=\dotu_z(\Point=\Point_L, t_i)$, i.e., the simulated velocity in $z$ direction at point $L$ (located at $\Point_L$) at time $t_i$ and
\item the voltage $V(t_i)$ measured by the PZ
sensor and the simulated quantity $\SimulatedPZSensorVoltageAtTmeas$, i.e., the simulated potential at the PZ sensor at time $t_i$
\end{enumerate}
for the discrete measurement times $t_i \in \MeasurementTimesSet \subseteq \TimeStepSet$ with $i\in\{1,2,\ldots, \NumMeasurements\}$, such that the set of measurement times $\MeasurementTimesSet$ is a subset of the discrete time steps $\TimeStepSet$ and where $\NumMeasurements$ is the number of measurements with $\NumMeasurements\leq\NumTimeSteps$.

We introduce an abstract form of the state problem (\ref{eq:dsas:a})--(\ref{eq:dsas:f}) (see
also \cite{Rohan_Cimrman_2002}), which can be written as
\begin{equation}
  \label{eq:asp}
  \Phi(\ParameterVector, \ybold^n, \ybold^{\PreviousTimestep}) = 0 \;, \quad \ybold^n \equiv [\NodalDisplacementsAtTcurrent, \NodalVelocitiesAtTcurrent,
  \NodalAccelerationsAtTcurrent, \NodalPotentialAtTcurrent, \NodalDotPotentialAtTcurrent, \SensorPotentialAtTcurrent]^T \;,
\end{equation}
where $\ParameterVector \in \R^{n_\Parameter}$ is a vector that collects the parameters to be identified,
and $\ybold^n$, $\ybold^{\PreviousTimestep}$ collect all states at time steps $n$ and $\PreviousTimestep$, respectively.

The objective of the parameter identification is to find the optimal parameter vector $\hat\ParameterVector$ defined as:
\begin{equation}
  \label{eq:identification}
  \hat\ParameterVector =
  \argmin_{%
    \substack{%
      \text{s.\,t.}\, \ParameterVector_{\min} \leq \ParameterVector  \leq \ParameterVector_{\max} \\
      \phantom{\text{s.\,t.}}\, \Phi(\ParameterVector, \ybold^n, \ybold^{\PreviousTimestep}) = 0
      \text{ for all time steps}
    }
  }
  F(\ParameterVector,\MeasurementData) \;,
\end{equation}
where $\ParameterVector_{\min}$ and $\ParameterVector_{\max}$ are the lower and upper bounds of the parameters, respectively.
The matrix of measurement data $\MeasurementData$ is defined as
\begin{equation}
   \MeasurementData = [\bar\ybold^1,\bar\ybold^2,\ldots,\bar\ybold^{\NumMeasurements}] \;,
\end{equation}
which collects the experimental data vectors obtained at time steps $t^i$ denoted by $\bar\ybold^i=[v_z(t_i), V(t^i)]^T$.
The objective function is of general least-squares form:
\begin{align}
  \label{eq:objective}
  F(\ParameterVector,\MeasurementData) &= F_{v_z}(\ParameterVector,\MeasurementData) + F_{V}(\ParameterVector,\MeasurementData) \\
  \label{eq:objective_detailed}
  &= \tfrac{1}{2}\sum_{i = 1}^{\NumMeasurements}\!\left(\tfrac{\SimulatedVibrometerVelocityAtTmeas - v_z(t_i)}{\NormalizationVelocity}\right)^2
  + \tfrac{1}{2}\sum_{i=1}^{\NumMeasurements}\!\left(\tfrac{\SimulatedPZSensorVoltageAtTmeas - V(t_i)}{\NormalizationVoltage}\right)^2 \;.
\end{align}
The objective function consists of two individual contributions,
the least-squares error in the velocity $F_{v_z}$ and the least-squares error in the voltage $F_V$.
Note that in \eqref{eq:objective_detailed}, the dependence of the objective function on the parameter vector $\ParameterVector$ is implicit, since evaluating it requires solving \eqref{eq:asp} to obtain the simulated quantities $\SimulatedVibrometerVelocityAtTmeas$ and $\SimulatedPZSensorVoltageAtTmeas$.
The individual summands of \eqref{eq:objective_detailed} are normalized by their respective experimental amplitudes to prevent either signal (mechanical velocity or electrical voltage) from dominating the objective function and to ensure a balanced contribution of both quantities to the optimization.
Correspondingly, the normalizing constants are defined as
\begin{equation}
  \NormalizationVelocity = \max_{i \in \{1, \ldots, n\}} |v_z(t_i)|,
  \qquad
  \NormalizationVoltage = \max_{i \in \{1, \ldots, n\}} |V(t_i)|.
  \label{eq:scaling_factors}
\end{equation}
Some algorithms for solving \eqref{eq:identification} require the gradient of the objective function with respect to the parameters, $\nabla_{\ParameterVector} F$.
Computing said gradient involves calculating the partial
derivatives $\pder{\ybold^n}{\Parameter_k}$, where $\Parameter_k$ is a component of $\ParameterVector$.
By differentiating (\ref{eq:asp}) w.r.t.
$\ParameterVector$, with $\Phi^n$ being a shorthand for $\Phi(\ParameterVector, \ybold^n, \ybold^{\PreviousTimestep})$, a recurrent relation
\begin{equation}
  \label{eq:dasp}
  \pder{\Phi^n}{\ybold^n} \pder{\ybold^n}{\ParameterVector} = - \left(\pder{\Phi^n}{\ParameterVector} +
  \pder{\Phi^n}{\ybold^{\PreviousTimestep}} \pder{\ybold^{\PreviousTimestep}}{\ParameterVector}\right)
\end{equation}
is obtained allowing us to compute $\pder{\ybold^n}{\ParameterVector}$ in all time steps,
initialized by using the initial conditions. Then
$\pder{\SimulatedVibrometerVelocityAtTmeas}{\Parameter_k}$ and $\pder{\SimulatedPZSensorVoltageAtTmeas}{\Parameter_k}$ are simply
components in $\pder{\ybold^n}{\ParameterVector}$.
This so-called \emph{direct differentiation method} requires solving, at each time step, a linear system with system matrix $\pder{\Phi^n}{\ybold^n}$ and $n_\Parameter$ right-hand sides. The blocks of this matrix involve the discrete system matrices introduced in \eqref{eq:dsas:a}–\eqref{eq:dsas:f}.

\section{Comparison of experiment and numerical modeling}

The aim of this section is twofold: first, to demonstrate the accuracy of the numerical model in reproducing the experiment described in Section~\ref{sec:experiments}, and second, to identify the proportional damping parameters, which are difficult to determine by modeling alone, as well as the parameters of the external electrical circuit.

\subsection{Initial setting of the material parameters}
\label{sec:ismp}

The density and elastic parameters of the steel beam were obtained from the measurements described in Section~\ref{sec:mmpb}. The parameters of the PZ disc corresponding to the PIC~181 sensor were provided by the manufacturer (PI Ceramic). The initial values of the proportional damping coefficients and the external circuit parameters were chosen ad hoc. The following material parameters were therefore used in the numerical simulations:
\begin{itemize}
\item Steel elastic beam: $\rho = 8014.5$~kg/m$^3$, $E = 189$~GPa, $\nu = 0.3$.
\item Piezoelectric disc: $\rho = 7890$~kg/m$^3$, vacuum permittivity
  $\epsilon^0 = 8.8541878128 \cdot 10^{-12}$~F/m and
\begin{equation}
  \begin{split}
    \mbox{in Voigt notation: } \Cb^P = \smatrix{rrrrrr}{[}
     144.1 &  79.65&  81.45&   0   &   0   &   0    \\
      79.65& 144.1 &  81.45&   0   &   0   &   0    \\
      81.45&  81.45& 134.4 &   0   &   0   &   0    \\
       0   &   0   &   0   &  27.29&   0   &   0    \\
       0   &   0   &   0   &   0   &  27.29&   0    \\
       0   &   0   &   0   &   0   &   0   &  32.22 \\
    \ematrix{]} \mbox{ GPa,} \\
    \eb = \smatrix{rrrrrr}{[}
 0 &  0 &  0 &  0 & 10.7 & 0 \\
 0 &  0 &  0 & 10.7 &  0 & 0 \\
-5.256 & -5.256 & 14.53 &  0 &  0 & 0 \\
    \ematrix{]} \mbox{ C/m$^2$,} \\
    \epssb = \epsilon^0 \smatrix{rrr}{[}
   717 &   0 &    0 \\
     0 & 717 &    0 \\
     0 &   0 & 665 \\
    \ematrix{]} \mbox{F/m} \;.
  \end{split}
\end{equation}
\item Proportional damping: none, i.e., $\alpha = 0$~\si{\per\s}, $\beta = 0$~\si{\s}.
\item Initial external circuit parameters: $R = 10$~\si{\mega\ohm},
  $C = 1$~\si{\nano\farad}.
\end{itemize}

The finite element model and the parameter identification procedure were implemented using the open-source finite element package SfePy \cite{Cimrman_Lukes_Rohan_2019, Cimrman_2021} in the Python programming language. The results of the direct simulations were additionally verified using COMSOL Multiphysics\textsuperscript{\textregistered}.
\subsection{Modal analysis of the beam without PZ disc}

A suitable mesh size for the subsequent SfePy simulations was determined through a modal analysis of the elastic beam. The lowest natural frequencies, computed for several mesh resolutions and displacement approximation orders 1 and 2, were compared with the analytical estimate $f_{1,,\mathrm{analytical}} = 143.62$~Hz (see Section~\ref{sec:analytical}) and with results obtained using COMSOL, where an analogous study was performed.

The mesh shown in Fig.~\ref{fig:domain}, considering only the beam part for this test, together with a quadratic displacement approximation, was found to provide satisfactory accuracy. Further mesh refinement or an increase in the approximation order did not lead to significant improvements. This choice also reflects our preference for using the coarsest possible mesh in the subsequent dynamic simulations of the experiments. For the selected mesh, the first eigenfrequency of the beam was computed as $f_1 = 145.44$~Hz. For comparison, the corresponding COMSOL result, obtained using a hexahedral mesh, was $f_1 = 145.31$~Hz.

The complete selected mesh consisted of 2121 vertices and 4815 (beam) + 1762 (PZ disc) tetrahedral elements. The corresponding COMSOL mesh contained 13,066 vertices and 8870 (beam) hexahedral elements, 690 (beam) prismatic elements, and 1190 (PZ disc) prismatic elements.

\subsection{Modal analysis of the beam with the PZ disc}

\begin{figure}[htp!]
\centering
\includegraphics[width=0.6\linewidth]{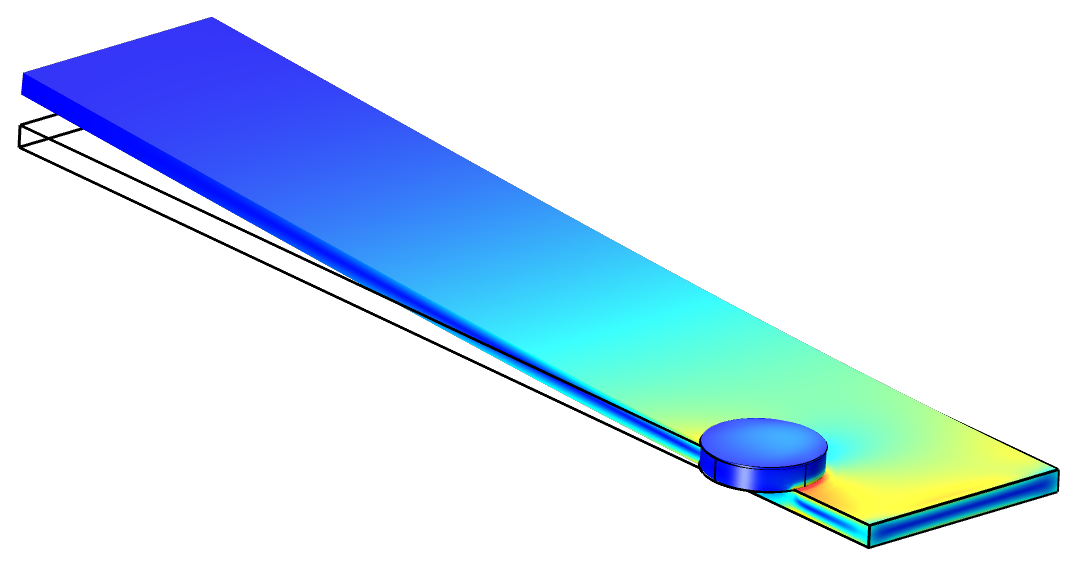}
\caption{First eigenmode computed with COMSOL corresponding to $f_1 = 150.57$~Hz; color indicates the von Mises stress.}
\label{fig:eigenmode}
\end{figure}

Subsequently, the complete selected mesh, comprising both the elastic beam and the PZ disc, was used for the modal analysis of the static part of \eqref{eq:wpedp:a} and \eqref{eq:wpedp:b}. The resulting first eigenfrequency obtained with SfePy was $f_1 = 150.74$~Hz, while the corresponding COMSOL result was $f_1 = 150.57$~Hz. The associated eigenmode is shown in Fig.~\ref{fig:eigenmode}. The close agreement between the computed frequencies confirms the consistency of the two discretizations. Consequently, the respective SfePy and COMSOL meshes were employed for the transient simulations and the parameter identification procedure described in the following sections.

The experimentally measured value, however, was lower, namely $f_{1,,\mathrm{measured}} = 147.59$~Hz (see Section~\ref{sec:oe}), as discussed further below.

\subsection{Transient problem}

The transient problem defined by \eqref{eq:wpedp:a}–\eqref{eq:wpedp:c} directly corresponds to the experiment described in Section~\ref{sec:problem}, and its finite element discretization is given by \eqref{eq:dsas:a}–\eqref{eq:dsas:e}.

\begin{figure}[htp!]
  \centering
  \begin{tabular}{cc}
    A) $u_3(t)$ at the laser measurement point~$L$.
    &
    B) $Q(t)$ on the $\Gamma_{pQ}$ and $\Gamma_{p0}$ electrode surfaces.
    \\
    \includegraphics[width=0.48\linewidth]{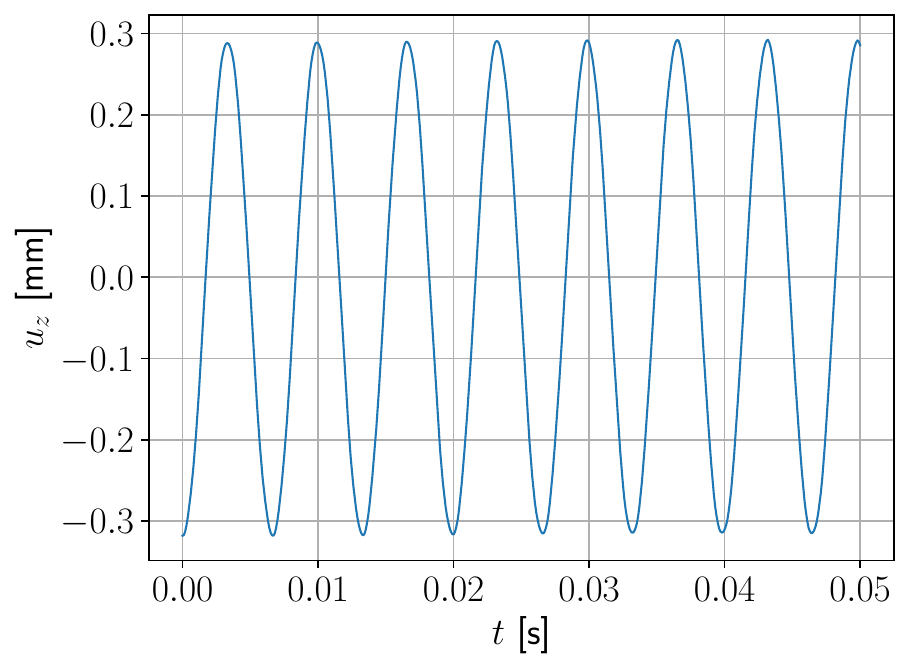}
    &
    \includegraphics[width=0.48\linewidth]{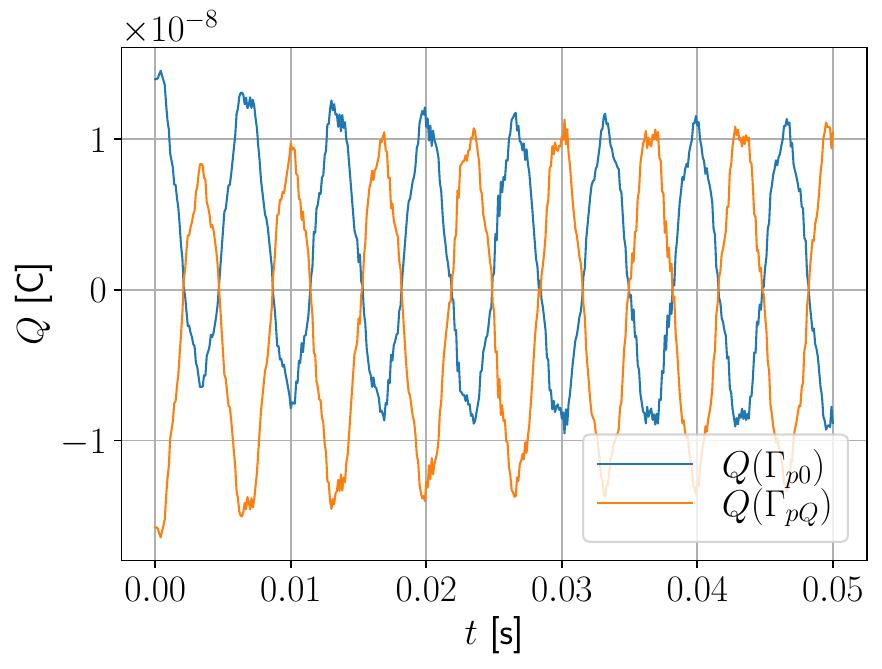}
    \\
    C) $\dot u_3(t)$ at the laser measurement point~$L$.
    &
    D) $\bar\vphi(t)$ on the $\Gamma_{pQ}$ electrode.
    \\
    \includegraphics[width=0.48\linewidth]{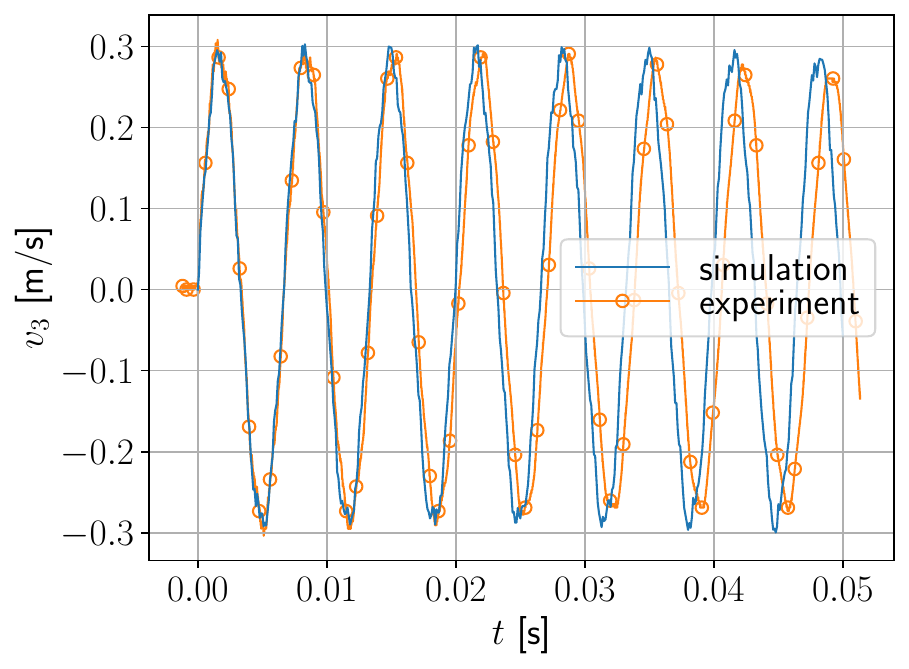}
    &
    \includegraphics[width=0.48\linewidth]{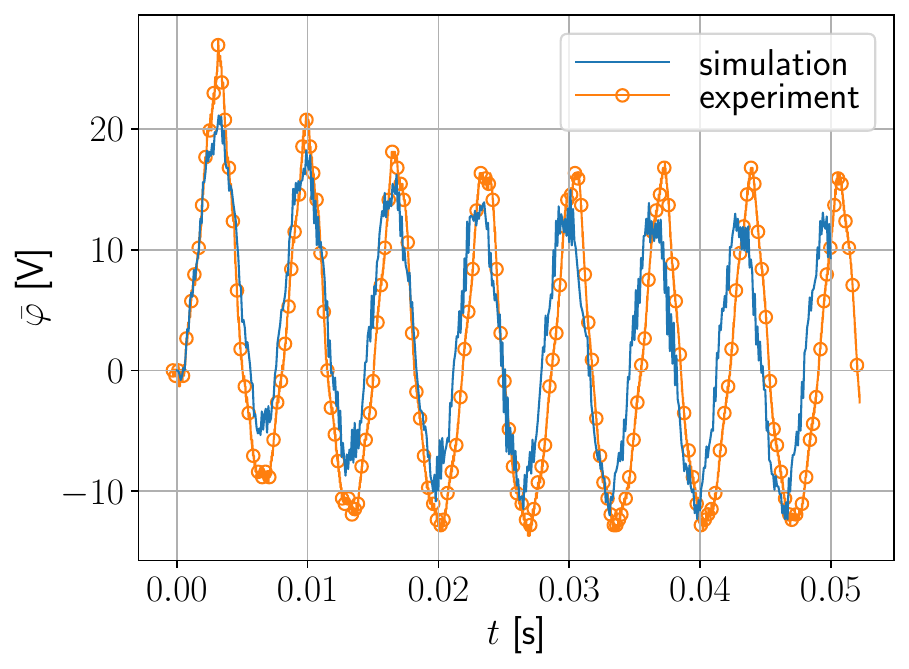}
  \end{tabular}
  \caption{Time histories of selected quantities for the initial material parameters. Comparison with experimental data.}
  \label{fig:transient-sfepy-initial}
\end{figure}

We first present results obtained using the initial parameter settings described in Section~\ref{sec:ismp}. Figure~\ref{fig:transient-sfepy-initial} shows the time histories of the following quantities:
\begin{itemize}
\item displacement $u_3(t)$ and velocity $\dot u_3(t)$ at the laser measurement point~$L$,
\item induced charges $Q(t)$ on the $\Gamma_{pQ}$ and $\Gamma_{p0}$ electrode surfaces of the sensor,
\item electric potential $\bar\vphi(t)$ on the $\Gamma_{pQ}$ electrode.
\end{itemize}
Comparison with the experimental data reveals a slight mismatch in the dominant frequency (see panel~C and the modal analysis results in the previous section), as well as discrepancies related to the external electrical circuit parameters (panel~D).

\begin{figure}[htp!]
  \centering
  \begin{tabular}{cc}
    A) $\dot u_3(t)$ at the laser measurement point~$L$.
    &
    B) $\bar\vphi(t)$ on the $\Gamma_{pQ}$ electrode.
    \\
    \includegraphics[width=0.48\linewidth]{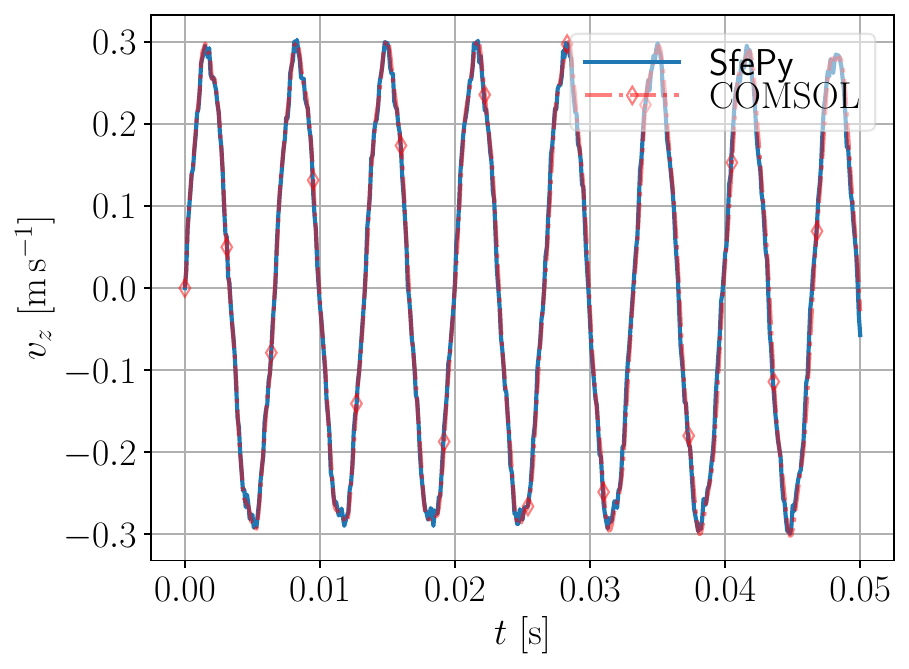}
    &
    \includegraphics[width=0.48\linewidth]{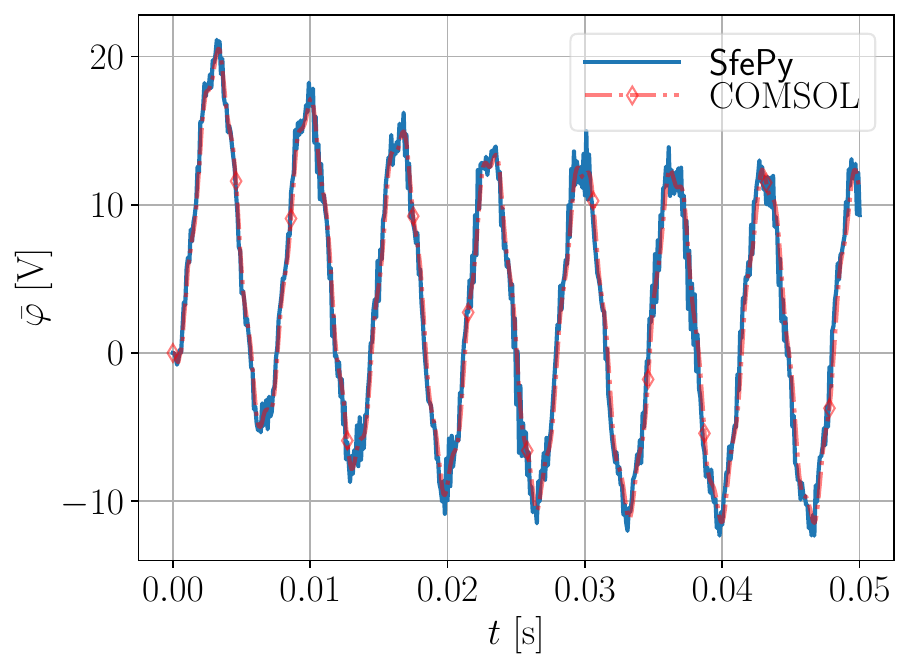}
  \end{tabular}
  \caption{Time histories of selected quantities for the initial material parameters. Comparison between SfePy and COMSOL results.}
  \label{fig:transient-sfepy-comsol-initial}
\end{figure}

Before adjusting the model parameters, the SfePy-based model was verified by comparing its transient response with results obtained using COMSOL. Despite the use of different meshes in the two models, very good agreement was observed, as shown in Fig.~\ref{fig:transient-sfepy-comsol-initial}. This confirms the correctness of the numerical implementation.

\begin{figure}[htp!]
  \centering
  \begin{tabular}{cc}
    A) $\dot u_3(t)$ at the laser measurement point~$L$.
    &
    B) $\bar\vphi(t)$ on the $\Gamma_{pQ}$ electrode.
    \\
    \includegraphics[width=0.48\linewidth]{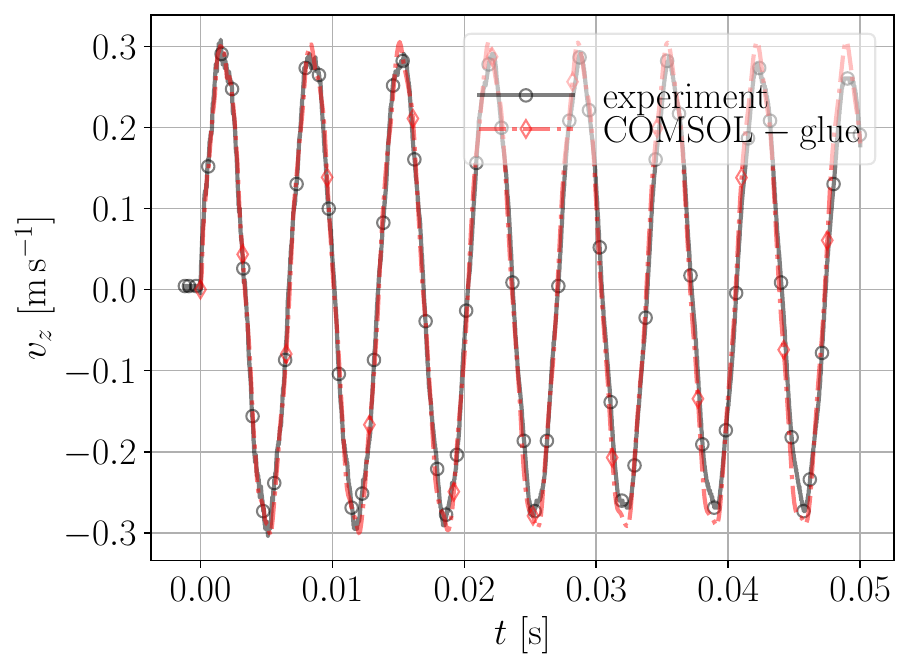}
    &
    \includegraphics[width=0.48\linewidth]{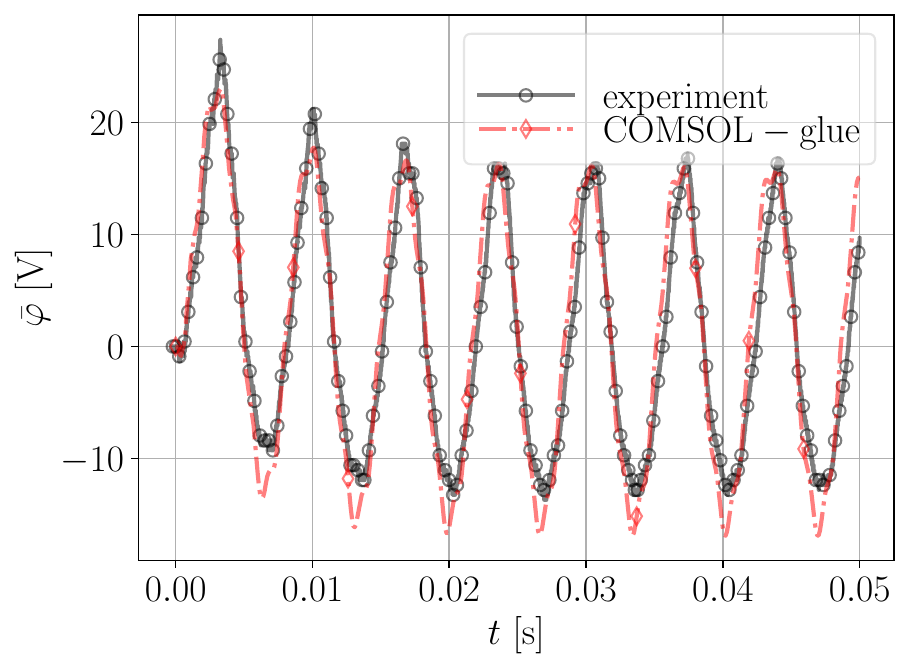}
  \end{tabular}
  \caption{Time histories of selected quantities for the initial material parameters. Comparison of experimental data with COMSOL results using the glue layer model and zero oscilloscope capacitance.}
  \label{fig:transient-comsol-glue-initial}
\end{figure}

The numerical model assumes a perfectly rigid attachment of the PZ sensor to the beam. One plausible explanation for the observed frequency discrepancy is that, in reality, the adhesive layer between the PZ disc and the beam exhibits finite elasticity. To investigate this effect, the COMSOL model was extended by introducing a thin glue layer with the following properties: Young’s modulus $3.4$~GPa, Poisson’s ratio $0.34$, thickness $0.2$~mm, and density $0~\si{kg/m^3}$. The resulting time histories, shown in Fig.~\ref{fig:transient-comsol-glue-initial}(A), demonstrate that this modification indeed reduces the principal frequency to a value closer to the experimentally observed one.

Furthermore, the computed electric potential closely matched the experimental data when the oscilloscope was modeled purely as a resistive element. Consequently, the oscilloscope capacitance was set to zero, in contrast to the initial parameter settings described in Section~\ref{sec:ismp}. Figure~\ref{fig:transient-comsol-glue-initial}(B) presents the resulting voltage time histories obtained using the glue layer model without oscilloscope capacitance, compared to the experimental measurements.

As an alternative approach, the principal frequency of the model can also be reduced by (artificially) decreasing the Young’s modulus of the beam. However, this approach retains the assumption of rigid mechanical coupling between the beam and the PZ disc, which leads to an overestimation of the generated voltage amplitude. This effect can be compensated by (artificially) increasing the oscilloscope capacitance, which is connected in parallel with the PZ disc. This strategy is explored in the next section, where the model parameters are identified automatically using the proposed parameter identification procedure.

\subsection{Identification of damping and electrical parameters}
To illustrate the parameter identification procedure, we determine the mechanical parameters $\alpha$ (mass-proportional damping), $\beta$ (stiffness-proportional damping), and $E$ (Young’s modulus of the beam), together with the electrical parameters $R$ (oscilloscope resistance) and $C$ (oscilloscope capacitance). The resulting parameter vector is $\ParameterVector = [\alpha, \beta, E, R, C]$.
Experimental vibrometer data $v_z(t_i)$ and voltage measurements $V(t_i)$ are available for $t_i \in [0, 0.05]$~s and are used for the parameter identification. In the following, two different parameter identification approaches are considered and compared.

\paragraph{Sequential parameter identification}
In the first approach, the mechanical and electrical parameters were identified sequentially using a least-squares solver. Specifically, we employed the \texttt{least\_squares()} function from SciPy \cite{scipy_2020} with the Trust Region Reflective (TRF) algorithm. This solver can either approximate the Jacobian of the objective function $F$ numerically using finite differences or utilize a user-supplied function providing the partial sensitivities.

For the present linear problem, no significant reduction in computational time is expected when using a semi-analytical Jacobian (see Section~\ref{sec:pi}) compared to the default two-point finite difference scheme, since both approaches require solving the same number of linear systems. Moreover, the analytical Jacobian entails additional computational overhead. Nevertheless, its implementation allowed us to test automatically differentiated terms based on JAX \cite{jax2023github} and serves as a proof of concept for future nonlinear extensions of the model.

Starting from the initial guesses and parameter bounds listed in Table~\ref{tab:results-identification}, the parameter values were identified as follows. First, only the mechanical parameters $\alpha$, $\beta$, and $E$ were identified by minimizing solely the velocity contribution $F_{v_z}$ in the objective function \eqref{eq:objective}, effectively setting the voltage contribution to zero, $F_V = 0$. During this step, the electrical parameters $R$ and $C$ were fixed at their initial values given in Table~\ref{tab:results-identification}. In a second least-squares run, the electrical parameters were identified by minimizing only the voltage contribution $F_V$, with $F_{v_z} = 0$, while keeping the mechanical parameters fixed at the values obtained in the first run. In both identification steps, the normalization factors $\NormalizationVelocity$ and $\NormalizationVoltage$ were set to unity, since scaling a single-term least-squares objective function by a constant does not affect the location of its minimum.

\begin{figure}[H]
  \centering
  \begin{tabular}{cc}
    A) Mechanical parameters.
    &
    B) Electrical parameters.
    \\
    \includegraphics[width=0.48\linewidth]{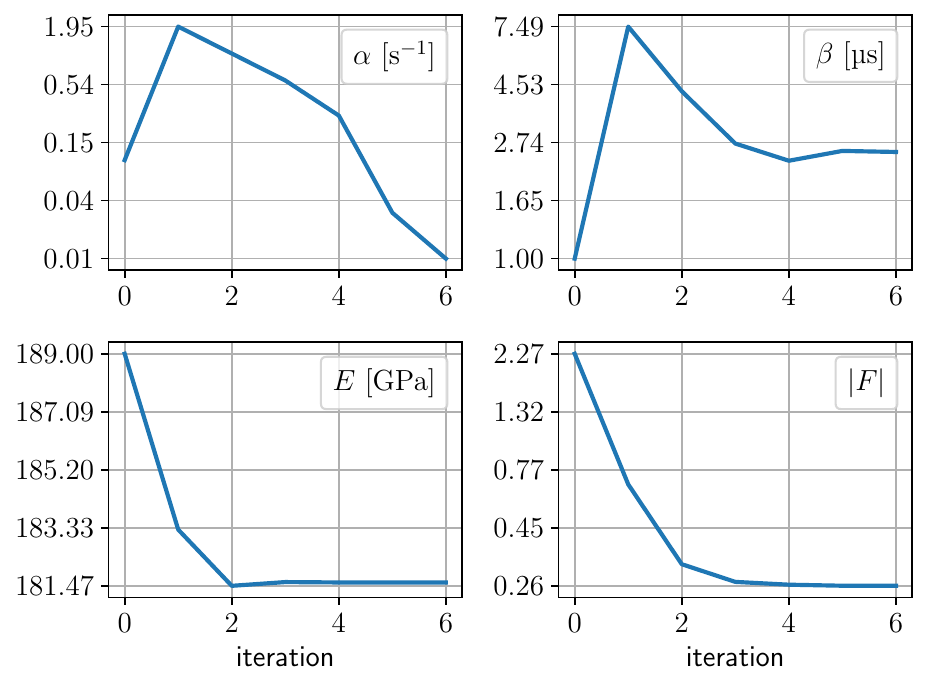}
    &
    \includegraphics[width=0.48\linewidth]{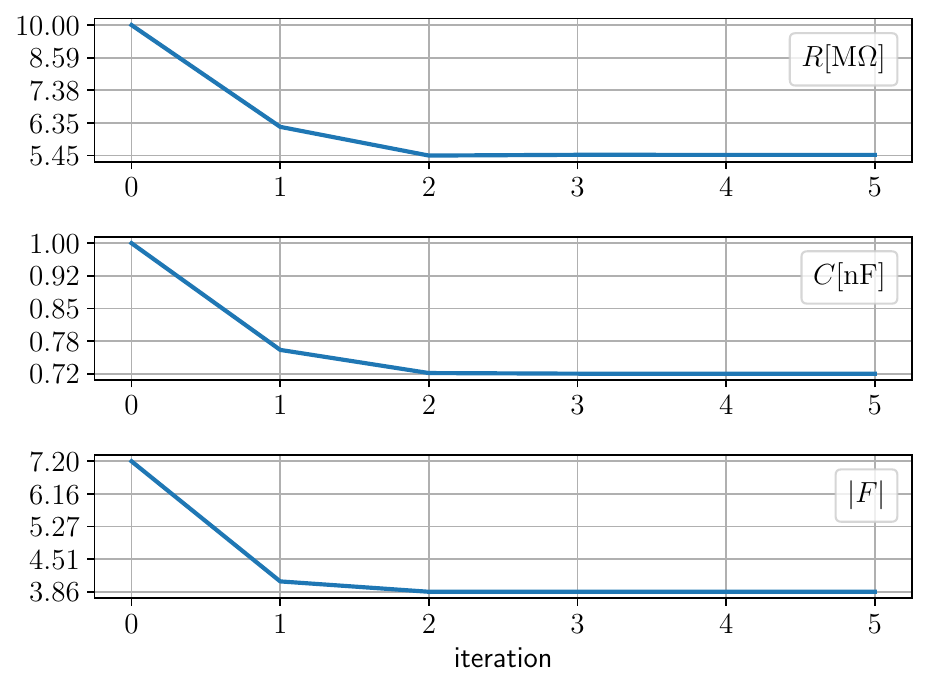}
  \end{tabular}
  \caption{Evolution of the material parameters and the objective function $|F|$ during the sequential identification using SciPy’s \texttt{least\_squares()} function with the TRF algorithm. Panels A) and B) correspond to separate identification runs: in the first run (A), the electrical parameters were fixed at their initial values, while in the second run (B) the mechanical parameters were fixed to those obtained in the first run, such that each step optimized only the respective subset of parameters.}
  \label{fig:pars-identification}
\end{figure}

\begin{figure}[H]
  \centering

  \begin{subfigure}[t]{0.48\linewidth}
    \centering
    \caption{$\dot u_3(t)$ at the laser measurement point~$L$}
    \label{fig:transient-u3-initial}
    \includegraphics[width=\linewidth]{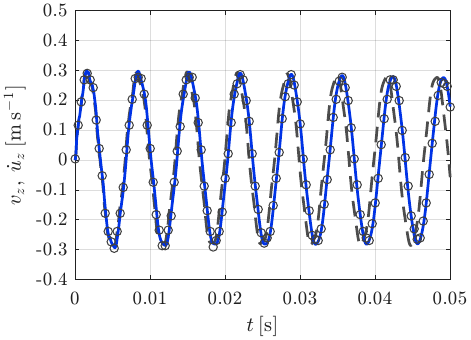}
  \end{subfigure}
  \hfill
  \begin{subfigure}[t]{0.48\linewidth}
    \centering
    \caption{$\bar\vphi(t)$ on the $\Gamma_{pQ}$ electrode}
    \label{fig:transient-phi-initial}
    \includegraphics[width=\linewidth]{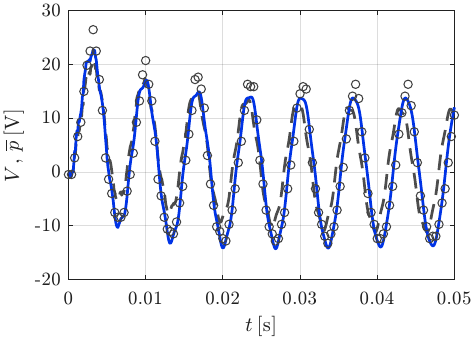}
  \end{subfigure}

  \vspace{3mm}

  \begin{subfigure}[t]{0.48\linewidth}
    \centering
    \caption{CMA-ES vs.\ experiment (velocity)}
    \label{fig:transient-u3-cmaes}
    \includegraphics[width=\linewidth]{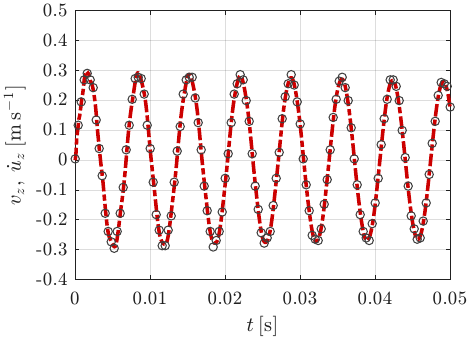}
  \end{subfigure}
  \hfill
  \begin{subfigure}[t]{0.48\linewidth}
    \centering
    \caption{CMA-ES vs.\ experiment (potential)}
    \label{fig:transient-phi-cmaes}
    \includegraphics[width=\linewidth]{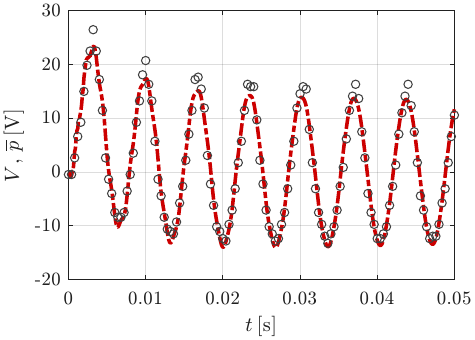}
  \end{subfigure}

  \vspace{3mm}

  \begin{subfigure}[t]{0.48\linewidth}
    \centering
    \caption{Difference in $\dot u_z$}
    \label{fig:transient-u3-diff}
    \includegraphics[width=\linewidth]{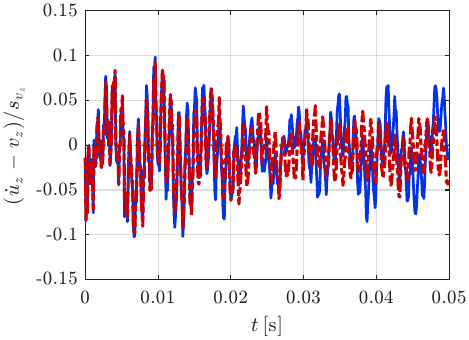}
  \end{subfigure}
  \hfill
  \begin{subfigure}[t]{0.48\linewidth}
    \centering
    \caption{Difference in $\bar\vphi$}
    \label{fig:transient-phi-diff}
    \includegraphics[width=\linewidth]{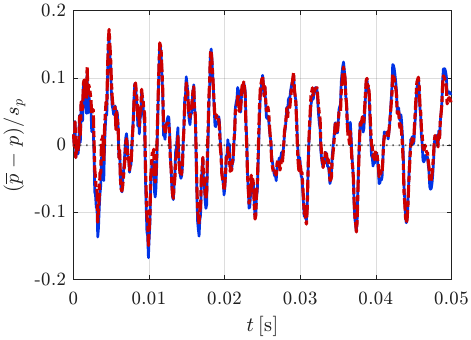}
  \end{subfigure}

  \vspace{3mm}

  \begin{subfigure}{0.6\linewidth}
    \centering
    \includegraphics[width=\linewidth]{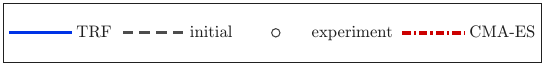}
  \end{subfigure}

  \caption{Time histories and normalized differences of selected quantities for the initial and identified material parameters, compared with the experimental data.}
  \label{fig:transient-sfepy-identification}
\end{figure}

The evolution of the parameters over the solver iterations is shown in Fig.~\ref{fig:pars-identification}, and the final identified values are summarized in Table~\ref{tab:results-identification}. The identified parameters yield a first eigenfrequency of $f_1 = 147.82$~Hz. The simulated time histories of $\SimulatedVibrometerVelocity(t)$ and $\SimulatedPZSensorVoltage(t)$ obtained using the initial and the identified parameters are compared with the experimental data in Fig.~\ref{fig:transient-u3-initial} and Fig.~\ref{fig:transient-phi-initial}, respectively.

\paragraph{Global parameter identification}
In the second parameter identification approach, a simultaneous identification of the mechanical and electrical parameters was carried out using the Covariance Matrix Adaptation Evolution Strategy (CMA-ES) \cite{Hansen2001}. We employed the implementation provided by the \texttt{pymoo} library \cite{pymoo} in combination with the QUEENS framework for solver-independent multi-query analyses of large-scale computational models \cite{queens}.
To avoid overfitting to high-frequency noise, only a representative subset of the experimental time series (as shown in Fig.~\ref{fig:transient-sfepy-identification}) was used during the optimization. Prior to evaluating the objective function, the simulated quantities $\SimulatedVibrometerVelocity(t)$ and $\SimulatedPZSensorVoltage(t)$ were linearly interpolated to the experimental measurement times $t_i$.

The motivation for this global optimization strategy was twofold: to capture potential coupling effects between the parameters within a single optimization run, and to ensure that the identified parameters correspond to a minimum of the full objective function \eqref{eq:objective}. The optimization was performed using a population size of 16 over a total of 400 generations. As in the first approach, the algorithm was initialized using the initial guesses and parameter bounds reported in Table~\ref{tab:results-identification}. The identified parameter values are summarized in Table~\ref{tab:results-identification}. The corresponding simulated time histories of $\SimulatedVibrometerVelocity(t)$ and $\SimulatedPZSensorVoltage(t)$ are compared with the experimental data in Fig.~\ref{fig:transient-u3-cmaes} and Fig.~\ref{fig:transient-phi-cmaes}, respectively.

A comparison of the two identification strategies reveals a substantial difference in computational cost. The sequential TRF-based approach required only 39 objective function evaluations, whereas the global CMA-ES method involved 6400 evaluations due to its population-based nature. It should be noted, however, that in the sequential approach the parameters are not identified within a single optimization run; instead, each subset of parameters is optimized separately. Despite this difference, both strategies yielded comparable parameter estimates. The final value of the full objective function was 0.36 for the sequential TRF-based identification and 0.32 for the global CMA-ES approach, indicating a slightly better fit for the latter while remaining within the same order of magnitude.


\begin{table*}[htp!]
  \centering
  \begin{tabular}{lrrrr}
    \toprule
    & initial & bounds & final (TRF) & final (CMA-ES) \\
    \midrule
    $\alpha$~[\si{\per\s}]         & 0.1  & [0.01,\,10]           & 0.011 & 0.2981 \\
    $\beta$~[\si{\micro\s}]        & 1    & [0.1,\,100]           & 2.5   & 4.791 \\
    $E$~[GPa]                      & 189  & [140,\,200]           & 182   & 181.46 \\
    $R$~[\si{\mega\ohm}]           & 10   & [1,\,50]              & 5.5   & 5.73 \\
    $C$~[\si{\nano\farad}]         & 1    & [0.01,\,2]            & 0.72  & 0.685 \\
    \midrule
    $F$          & 9.85 & ---                   & 0.36  & 0.32 \\
    \bottomrule
  \end{tabular}
  \caption{Initial guesses, bounds $[\ParameterVector_{\min}, \ParameterVector_{\max}]$, and optimal parameter values obtained using the TRF and CMA-ES algorithms—corresponding to the first and second identification approaches, respectively—are reported. The associated values of the full objective function $F$ (see \eqref{eq:objective}) are also listed. For the sequential identification based on the TRF algorithm, the reported value of $F$ represents the sum of the objective function values from the individual identification runs.}
  \label{tab:results-identification}
\end{table*}

\section{Conclusions}

In this paper, we presented experimental and numerical investigations of the dynamic response of a system consisting of a cantilever beam and a piezoelectric disc sensor. The system vibrations and the sensor voltage were measured using a custom-built experimental setup. The first eigenfrequency was extracted from the measured data and compared with the analytical value obtained for the beam alone, showing good agreement. The remaining discrepancy can be attributed to damping effects in the experiment, as the eigenfrequency of a damped system is lower than that of the corresponding undamped system (see \cite{rao1995mechanical}).
The complete experimental configuration was also modeled mathematically, with spatial discretization performed using the finite element method and temporal discretization using a second-order time integration scheme. Since several model parameters, such as the proportional damping coefficients of the Rayleigh model, cannot be measured directly, selected parameters were identified using two different approaches: a gradient-based Trust Region Reflective (TRF) algorithm and the gradient-free Covariance Matrix Adaptation Evolution Strategy (\mbox{CMA-ES}). The identification results indicate that the TRF-based approach provides a computationally efficient means of achieving a satisfactory fit, while the \mbox{CMA-ES} method yields slightly improved accuracy at the expense of a higher computational cost. Owing to its well-controlled experimental setup and comprehensive numerical treatment, the presented experiment can serve as a useful benchmark for numerical modeling and parameter identification in piezo-elastodynamic systems.
\paragraph{Acknowledgements}
{
The research was supported by the Czech Science Foundation (CSF) under grant No.~22-00863K within the institutional support RVO:61388998. The authors also acknowledge V\'{a}clav Kolman for the design of the experimental holder.
Additional funding was provided by the Deutsche Forschungsgemeinschaft (DFG, German Research Foundation) under grant number~490743767.
}

\bibliography{piezo-identification.bib}

\end{document}